# An application of the theta operator in generalized hypergeometric coherent states formalism

**Dušan POPOV**
University Politehnica Timisoara, Romania
Department of Physical Foundations of Engineering
Foreign member of Serbian Academy of Nonlinear Sciences (SANS), Belgrade, Serbia
E-mail: dusan_popov@yahoo.co.uk
ORCID: https://orcid.org/0000-0003-3631-3247

**Abstract**

In this paper we examine one of the multiple applications of the theta operator *xd/dx* in quantum mechanics, namely, in the formalism of generalized hypergeometric coherent states (GHG CSs). These states are the most general coherent states, in the sense that from them, through particularization, all coherent states with physical meaning can be obtained. A series of new results were obtained and some already known ones were found / confirmed (the integral representations, as well as the Laplace transform of hypergeometric functions). To support the theoretical considerations presented above, we examined, as example, the quantum systems with a linear energy spectrum. The results obtained in this paper contribute to widening the area of applicability of the theta operator.

**1. Introduction**

From a historical point of view, it is known that the operator $\vartheta_x \equiv x\dfrac{\partial}{\partial x} = x\,\partial_x$ , sometimes also called *theta operator* [1], has been used since the time of Euler, who apparently used it for the first time, in 1740, for the study of the convergence of power series [2]. It proved useful in the calculation of the sum of power series, respectively in the study of the properties of arbitrary infinitely differentiable functions. Later, the theta operator aroused more interest in other aspects, so that Hayden took up the study of the inverse theta operator [3]:

$$\vartheta_x \left(\vartheta_x\right)^{-1} = \left(\vartheta_x\right)^{-1}\vartheta_x = 1 \quad , \quad \left(\vartheta_x\right)^{-1} = \left(\dfrac{\partial}{\partial x}\right)^{-1} x^{-1} \tag{1.01}$$

In the last decades, it has been proven that the operator $\vartheta_x$ has multiple applications not only in the study of the properties of power series or generalized polynomials [2], [4], [5], but also in theoretical physics, specifically in quantum mechanics. In quantum mechanics, where orthogonal polynomials or hypergeometric functions intervene, as the solutions of Schrodinger's equation for different quantum systems, the theta operator $\vartheta_x$ definitively entered the mathematical theory of quantum mechanics.

On the other hand, the notion of coherent states was introduced by Schrodinger in 1926 [6]. This referred to the linear harmonic oscillator (HO-1D), being defined as the quantum states with the greatest similarity, as well as temporal behavior, to the classical states. However, Schrodinger's notion of coherent states did not immediately arouse much scientific interest. The revitalization of this notion was achieved in the second half of the last century, when coherent states began to be built for other quantum systems, different from HO-1D. The formulation of coherent states approach for more complicated quantum systems involved the introduction and use of generalized hypergeometric functions, as normalization functions of new defined coherent states.

The pioneers of this development of the formalism of coherent states (now called "non-linear", as opposed to "linear", corresponding to HO-1D), were, through their works: J.R. Klauder and B. Skagerstam [7], A. Perelomov [8], W-M. Zhang, D. H. Feng, and R. Gilmore [9], J-P. Gazeau [10], for more references see, e.g. the excellent review-paper of V.V. Dodonov [11].

## 2. Some remarkable properties of the $\vartheta_x$ operator

First, we will specify that, in order to simplify the writing of the formulas, we will use the notation for theta operator: $\vartheta_x \equiv x \frac{\partial}{\partial x}$, in which, by the variable $x$, we will understand the quantities (real or complex variables, or operators): $x = \{z^*, z^*z', |z|^2, \mathcal{A}_+, \mathcal{A}_-, \mathcal{A}_+\mathcal{A}_-\}$, depending on the case.

In most of the cases that we will analyze, the operator $\vartheta_x$ will act either on a power of $x$, i.e. $x^n$, or on a function of $f(x)$. That is why it is useful to recall some of the properties of the $\vartheta_x$ operator [2], [4].

The main property or conclusion is the fact that $x^n$ is an eigenfunction of the theta operator $\vartheta_x \equiv x \frac{\partial}{\partial x}$ to eigenvalue $n$. For this reason it is sometimes also called the homogeneity operator, because its eigenfunctions are the monomials in $x$:

$$\vartheta_x x^n = n\, x^n \tag{2.01}$$

Starting from this relation, we obtain some interesting results:

$$(\vartheta_x)^m = n^m\, x^n \tag{2.02}$$

The commutators are

$$[\vartheta_x,\, x] = x \quad,\quad [(\vartheta_x)^m,\, x^n] = n^m\, x^n \tag{2.03}$$

$$[(\vartheta_x)^m,\, \hat{f}(x)] = (\vartheta_x)^m\, \hat{f}(x) \tag{2.04}$$

Other relations

$$\hat{f}(\vartheta_x)g(x) = \sum_{n=0}^{\infty} g_n \, f(n) x^n \tag{2.05}$$

$$\hat{f}(\vartheta_x)\exp\left[-\beta \hat{f}(\vartheta_x)\right] = \exp\left[-\beta \hat{f}(\vartheta_x)\right]\hat{f}(\vartheta_x) \tag{2.06}$$

$$\exp(\gamma \vartheta_x)x^n = \exp(\gamma n)x^n \quad , \quad \gamma = \text{const} \tag{2.07}$$

$$\exp\left[\gamma \hat{f}(\vartheta_x)\right]x^n = \exp\left[\gamma f(n)\right]x^n \tag{2.08}$$

$$\exp\left[\gamma \hat{f}(\vartheta_x)\right]g(x) = f(e^\gamma x) \tag{2.09}$$

From these relations, the following formal substitution can be observed

$$n \Leftrightarrow \vartheta_x = x\frac{\partial}{\partial x} \quad , \quad f(n) \Leftrightarrow \hat{f}(\vartheta_x) \tag{2.10}$$

that is, in various expressions, where it is required, we can replace the principal quantum number $n$ with the operator $\vartheta_x = x\frac{\partial}{\partial x}$.

Assuming that the eigenvalue equation of the Hamiltonian of the examined system is

$$\hat{\mathcal{H}}|n> = \hbar\omega\, e(n)|n> \tag{2.11}$$

where $e(n)$ are the dimensionless energy eigenvalues, we obtain another series of interesting results, which will be useful in the following:

$$\exp\left[-\beta\hbar\omega\, e(\vartheta_x)\right]\,_pF_q(a;b;x) = \,_pF_q(a;b;e^{-\beta\hbar\omega}x) \tag{2.12}$$

where $_pF_q(a;b;x)$ is the generalized hypergeometric (GH) function.

On the other hand, let's also recall the familiar expansion of theta operator acting on an arbitrary infinitely differentiable function

$$\left(x\frac{\partial}{\partial x}\right)^n f(x) = \sum_{k=0}^{n} S(n,k) x^k \frac{\partial^k}{\partial x^k} f(x) \tag{2.11}$$

in which Stirling's numbers of the second kind $S(n,k)$ appear.

### 3. Generalized hypergeometric coherent states

Generally, the coherent states (CSs) are a set of ket vectors labeled by a complex number $z = |z|\exp(i\varphi)$, with $|z| \leq \mathcal{R}_c \leq \infty$, $0 \leq \varphi \leq 2\pi$, where $\mathcal{R}_c$ is the convergence radius. This set of

vectors can be expanded into some orthogonal basis, for example the Fock's vectors basis $\{|n>,\ n=0,1,2,...,n_{max}\leq\infty\}$:

$$|z> = \frac{1}{\sqrt{\mathcal{N}(|z|^2)}} \sum_n \frac{z^n}{\sqrt{\rho(n)}} |n> \qquad (3.01)$$

Here $\rho(n)$ are positive numbers which essentially determine the internal structure of CSs, and for this reason $\rho(n)$ are called the *structure constants*.

The CSs exists, on the one hand, if and only if the normalization function $\mathcal{N}(|z|^2)$ is expressed through an analytical function of real variable $|z|^2$. On the other hand, any CSs must fulfill a number of criteria summarized by Klauder (often called "*Klauder's prescriptions*"): continuity in complex label, normalization, non-orthogonality, resolution of unity operator with unique positive weight function of the integration measure, and supplementary, only for the CSs corresponding to the systems with infinite energy spectrum, the temporal stability and action identity [12].

The most general class of CSs are the *generalized hypergeometric coherent states* (GHG-CSs) whose appellation becomes from their normalization function which is given by a generalized hypergeometric function, e.g. $\mathcal{N}(|z|^2) = {}_pF_q(\boldsymbol{a};\boldsymbol{b};|z|^2)$. These kinds of states were firstly introduced by Appl and Schiller [13] and applied to the mixed (thermal) states in one of our previous papers [14]. Their expansion in the Fock-vectors basis is

$$|z> = \frac{1}{\sqrt{{}_pF_q(\boldsymbol{a};\boldsymbol{b};|z|^2)}} \sum_{n=0}^{\infty} \frac{z^n}{\sqrt{\rho_{p,q}(\boldsymbol{b}/\boldsymbol{a}\,|\,n)}} |n> \qquad (3.02)$$

The short notation used for the sequence of real numbers is $\{a_1, a_2, ..., a_p\} \equiv \{a_i\}_{i=1}^p \equiv \boldsymbol{a}$ and so on. By particularizing in a suitable manner these numbers, as well as the positive integers $p$ and $q$, we obtain all known CSs which have the physical meaning [15]. This is the reason why GHG-CSs are considered to be the *most general CSs*.

The generalized hypergeometric function which plays the role of normalization function is defined as [15]

$${}_pF_q(\boldsymbol{a};\boldsymbol{b};x) = \sum_{n=0}^{\infty} \frac{\prod_{i=1}^{p}(a_i)_n}{\prod_{j=1}^{q}(b_j)_n} \frac{x^n}{n!} \equiv \sum_{n=0}^{\infty} \frac{1}{\rho_{p,q}(\boldsymbol{b}/\boldsymbol{a}\,|\,n)} x^n \qquad (3.03)$$

In the definition, Pohhammer's symbols $(a)_n$ appear, defined with the help of Euler's gamma functions $\Gamma(a)$:

$$(a)_n = \frac{\Gamma(a+n)}{\Gamma(a)} = (a+n-1)(a+n-2)...(a+1)a. \qquad (3.04)$$

Also, some useful properties of Pochhammer's symbol are

$$\frac{(a)_n}{(a+n-1)} = (a)_{n-1} \qquad (3.05)$$

$$(a)_{n+1} = (a+n)(a)_n \quad , \quad (a+1)_n = \frac{1}{a}(a+n)(a)_n = \frac{1}{a}(a)_{n+1} \tag{3.06}$$

The generalized hypergeometric function can be represented through Meijer's G-function, as follows

$$_pF_q(a;b;x) = \frac{\prod_{j=1}^{q}\Gamma(b_j)}{\prod_{i=1}^{p}\Gamma(a_i)} G_{p,q+1}^{1,p}\left(-x \left| \begin{array}{cc} 1-a & ; & / \\ 0 & ; & 1-b \end{array} \right. \right) \tag{3.07}$$

Let's choose now a pair of Hermitical creation $\hat{\mathcal{A}}_+$ and annihilation $\hat{\mathcal{A}}_-$ operators, whose actions on the Fock vectors are [16]:

$$\hat{\mathcal{A}}_-|n> = \sqrt{e(n)}|n-1> \quad , \quad <n|\hat{\mathcal{A}}_+ = \sqrt{e(n+1)}<n+1| \tag{3.08}$$

We can impose that $e(n)$ be the (dimensionless) eigenvalues of the Hamiltonian's operator:

$$<n|\hat{\mathcal{A}}_+\hat{\mathcal{A}}_-|n> = e(n) \equiv <n|\hat{\mathcal{H}}|n> \tag{3.09}$$

These operators act on the vacuum state (state without particles) $|0>$ as follows:

$$|n> = \frac{1}{\sqrt{\rho_{p,q}(b/a|n)}}(\hat{\mathcal{A}}_+)^n|0> \quad , \quad <n| = \frac{1}{\sqrt{\rho_{p,q}(b/a|n)}}<0|(\hat{\mathcal{A}}_-)^n \tag{3.10}$$

The completeness relation of the Fock vectors leads to

$$\sum_{n=0}^{\infty}|n><n| = \sum_{n=0}^{\infty}\frac{1}{\rho_{p,q}(b/a|n)}(\hat{\mathcal{A}}_+)^n|0><0|(\hat{\mathcal{A}}_-)^n = 1 \tag{3.11}$$

In 2015 we introduced a new approach of normal ordering operator's products connected with the generalized hypergeometric CSs, called the *diagonal ordering operation technique* (DOOT) and denoted it with the symbol # # [16], as a generalization of another ordering technique - *the integration within an ordered product* (IWOP), introduced by Hong-yi Fan (see, e.g. [17] and references therein). But, the IWOP is applicable *only for Bose operators*, referring to the CSs of the HO-1D, while our ordering technique, DOOT is applied *to any pair* of raising and lowering nonlinear operators $\hat{\mathcal{A}}_+$ and $\hat{\mathcal{A}}_-$. In this sense, the IWOP is a particular case of DOOT.

The rules of the DOOT are: a) the order of operators $\hat{\mathcal{A}}_+$ and $\hat{\mathcal{A}}_-$ can be permuted inside the symbol # #, so that finally we obtain a function of normally ordered operator product $\#f(\hat{\mathcal{A}}_-\hat{\mathcal{A}}_+)\# = f(\hat{\mathcal{A}}_+\hat{\mathcal{A}}_-)$; b) inside the symbol # # we can perform all algebraic operations, according to the usual rules; c) the operators $\hat{\mathcal{A}}_+$ and $\hat{\mathcal{A}}_-$ are treated as simple *c*-numbers; d) the vacuum state projector $|0><0|$, in the frame of DOOT, has the following normal ordered form:

$$|0><0|= \frac{1}{\sum_{n=0}^{\infty}\frac{1}{\rho_{p,q}(b/a|n)}\#(\hat{\mathcal{A}}_+\hat{\mathcal{A}}_-)^n\#} = \frac{1}{\#_pF_q(a;b;\hat{\mathcal{A}}_+\hat{\mathcal{A}}_-)\#} \qquad (3.12)$$

We observe that in the denominator appear a normal ordered operator function depending on the ordered operator product $\#\hat{\mathcal{A}}_+\hat{\mathcal{A}}_-\#$ as "argument", as a consequence of the sentence c) according to which the operators $\hat{\mathcal{A}}_+$ and $\hat{\mathcal{A}}_-$ are treated as simple *c*-numbers.

It is easy to verify that, applying the DOOT rules on the completeness relation of the Fock vectors $\sum_{n=0}^{\infty}|n><n|=1$, the expression of the vacuum projector is obtained, as being the inverse function of the generalized hypergeometric function $\#_pF_q(a;b;\hat{\mathcal{A}}_+\hat{\mathcal{A}}_-)\#$. In the continuation of the paper, we will make full use of the fact that the ordered pair of operators $\#\hat{\mathcal{A}}_+\hat{\mathcal{A}}_-\#$ can be regarded as c-numbers, within the DOOT technique.

*Important note:* In the following, in order not to load the formulas, we will stop writing the signs #...# that indicate the normal ordering of the operators $\hat{\mathcal{A}}_+$ and $\hat{\mathcal{A}}_-$, i.e. $\hat{\mathcal{A}}_+\hat{\mathcal{A}}_-$ (the creation operator $\hat{\mathcal{A}}_+$ on the left, and the annihilation operator $\hat{\mathcal{A}}_-$ on the right) but *we will take into account that only this ordering exists.*

The generalized hypergeometric coherent states (GH-CSs) are obtained for choosing eigenvalues in the following form:

$$e(m) = m\frac{\prod_{j=1}^{q}(b_j+m-1)}{\prod_{i=1}^{p}(a_i+m-1)} \quad , \quad m=1,2,...,n \qquad (3.13)$$

to which the following structure function corresponds:

$$\rho_{p,q}(b/a|n) \equiv \prod_{m=1}^{n}e(m) = n!\frac{\prod_{j=1}^{q}(b_j)_n}{\prod_{i=1}^{p}(a_i)_n} \quad , \qquad (3.14)$$

and which leads to the normalization function $_pF_q(a;b;|z|^2)$.

*Note*: This choice of structure function corresponds to the Barut-Girardello type of coherent states. The procedure is similar for obtaining coherent states of the Klauder-Perelomov type, whose normalization function is $_qF_p(b;a;|z|^2)$, considering the dualism of the two types of states (for details related to this dualism, see [18]).

With these elements, GH CSs can be expressed as follows

$$|z> = \frac{1}{\sqrt{_pF_q(a;b;|z|^2)}}\sum_{n=0}^{\infty}\frac{z^n}{\sqrt{\rho_{p,q}(b/a|n)}}|n> = \frac{1}{\sqrt{_pF_q(a;b;|z|^2)}}\,_pF_q(a;b;z\hat{\mathcal{A}}_+)|0> \qquad (3.15)$$

and their complex cojugate

$$<z^*|=\frac{1}{\sqrt{_pF_q(a;b;|z|^2)}}\sum_{n=0}^{\infty}\frac{(z^*)^n}{\sqrt{\rho_{p,q}(b/a|n)}}<n|=\frac{1}{\sqrt{_pF_q(a;b;|z|^2)}}<0|\,_pF_q(a;b;z^*\hat{\mathcal{A}}_-) \quad (3.16)$$

so that the CSs projection operator is

$$|z><z^*|=\frac{1}{_pF_q(a;b;|z|^2)}\frac{_pF_q(a;b;z\hat{\mathcal{A}}_+)\,_pF_q(a;b;z^*\hat{\mathcal{A}}_-)}{_pF_q(a;b;\hat{\mathcal{A}}_+\hat{\mathcal{A}}_-)} \quad (3.17)$$

From the completeness relation of the GH CSs

$$\int d\mu_{p,q}(z)\,|z><z^*|=\sum_n|n><n|=1 \quad (3.18)$$

The weight function of the integration measure must be a unique positive defined function. To derive the integration measure $d\mu_{p,q}(z)=\frac{d\varphi}{2\pi}d(|z|^2)\,_ph_q(|z|)$, more precisely its weighting function $_ph_q(|z|)$, the procedure is as follows:

- We replace the CSs projector in the completeness relation:

$$\frac{1}{_pF_q(a;b;\hat{\mathcal{A}}_+\hat{\mathcal{A}}_-)}\int_0^{R_c}d(|z|^2)\frac{_ph_q(|z|)}{_pF_q(a;b;|z|^2)}\int_0^{2\pi}\frac{d\varphi}{2\pi}\,_pF_q(a;b;z\hat{\mathcal{A}}_+)\,_pF_q(a;b;z^*\hat{\mathcal{A}}_-)=1 \quad (3.19)$$

- We perform the following function change:

$$_p\tilde{h}_q(|z|)\equiv\frac{_ph_q(|z|)}{_pF_q(a;b;|z|^2)} \quad (3.20)$$

- We calculate the angular integral by applying the DOOT rules:

$$\int_0^{2\pi}\frac{d\varphi}{2\pi}\,_pF_q(a;b;z\hat{\mathcal{A}}_+)\,_pF_q(a;b;z^*\hat{\mathcal{A}}_-)=\sum_{n=0}^{\infty}\frac{(A_+A_-)^n}{[\rho_{p,q}(b/a|n)]^2}(|z|^2)^n \quad (3.21)$$

- After substitution, we have to solve the equation

$$\sum_{n=0}^{\infty}\frac{(A_+A_-)^n}{[\rho_{p,q}(b/a|n)]^2}\int_0^{R_c}d(|z|^2)\,_p\tilde{h}_q(|z|)(|z|^2)^n=\,_pF_q(a;b;\hat{\mathcal{A}}_+\hat{\mathcal{A}}_-) \quad (3.22)$$

- It can be seen that the condition is

$$\int_0^{R_c}d(|z|^2)\,_p\tilde{h}_q(|z|)(|z|^2)^n=\rho_{p,q}(b/a|n)=n!\frac{\prod_{j=1}^{q}(b_j)_n}{\prod_{i=1}^{p}(a_i)_n}=\frac{\prod_{i=1}^{p}\Gamma(a_i)}{\prod_{j=1}^{q}\Gamma(b_j)}\Gamma(n+1)\frac{\prod_{j=1}^{q}\Gamma(b_j+n)}{\prod_{i=1}^{p}\Gamma(a_i+n)} \quad (3.23)$$

- We change the exponent $s=n-1$.

$$\int_0^{R_c} d(|z|^2)\, _p\tilde{h}_q(|z|)\,(|z|^2)^{s-1} = \rho_{p,q}(b/a\,|\,n) == \frac{\prod_{i=1}^{p}\Gamma(a_i)}{\prod_{j=1}^{q}\Gamma(b_j)}\Gamma(s)\frac{\prod_{j=1}^{q}\Gamma(b_j+s-1)}{\prod_{i=1}^{p}\Gamma(a_i+s-1)}. \tag{3.24}$$

- The convergence radius of the GH-CSs $\mathcal{R}_c$ is the convergence radius of the hypergeometric series, given by [19]

$$\mathcal{R}_c = \liminf_{n\to\infty}\sqrt[n]{\rho(n)} = \lim_{n\to\infty}\frac{\rho(n)}{\rho(n+1)} = \begin{cases}\infty & \text{Stieltjes moment problem}\\ <\infty & \text{Hausdorff moment problem}\end{cases}. \tag{3.25}$$

- The solution of this integral equation is expressed through Meijer's G-functions [20]:

$$_p\tilde{h}_q(|z|) = \frac{\prod_{i=1}^{p}\Gamma(a_i)}{\prod_{j=1}^{q}\Gamma(b_j)} G_{p,q+1}^{q+1,0}\left(|z|^2 \,\middle|\, \begin{array}{c} /\ ;\quad \boldsymbol{a-1} \\ 0,\ \boldsymbol{b-1}\ ;\quad / \end{array}\right) \tag{3.26}$$

- Finally, the integration measure is [16]

$$d\mu(z) = \frac{\prod_{i=1}^{p}\Gamma(a_i)}{\prod_{j=1}^{q}\Gamma(b_j)}\frac{d\varphi}{2\pi}d(|z|^2)G_{p,q+1}^{q+1,0}\left(|z|^2 \,\middle|\, \begin{array}{c} /\ ;\quad \boldsymbol{a-1} \\ 0,\ \boldsymbol{b-1}\ ;\quad / \end{array}\right){}_pF_q(\boldsymbol{a};\boldsymbol{b};|z|^2) \tag{3.27}$$

Substituting this result in the expression of the completeness relation for GH-CSs, let's focus on the following integral in real space, which will play an important role in the entire GH-CSs $\Leftrightarrow$ DOOT $\Leftrightarrow \vartheta_x = x\dfrac{\partial}{\partial x}$ formalism:

$$\int_0^{R_c \leq \infty} d(|z|^2) G_{p,q+1}^{q+1,0}\left(|z|^2 \,\middle|\, \begin{array}{c} /\ ;\quad \boldsymbol{a-1} \\ 0,\ \boldsymbol{b-1}\ ;\quad / \end{array}\right)(|z|^2)^n = \frac{\prod_{j=1}^{q}\Gamma(b_j)}{\prod_{i=1}^{p}\Gamma(a_i)}\rho_{p,q}(b/a\,|\,n) \tag{3.28}$$

Generally, if $R_c \to \infty$, this relation is the Mellin transform of Meijer's G-functions [15]. Consequently, the completeness relation will be written, as an integral in the complex space

$$\int \frac{d^2z}{\pi} G_{p,q+1}^{q+1,0}\!\left(|z|^2 \left| \begin{array}{cc} / \; ; & a-1 \\ 0, \; b-1 \; ; & / \end{array} \right.\right) {}_pF_q(a;b;z\hat{\mathcal{A}}_+) \, {}_pF_q(a;b;z^*\hat{\mathcal{A}}_-) =$$

$$= \frac{\prod_{j=1}^{q} \Gamma(b_j)}{\prod_{i=1}^{p} \Gamma(a_i)} \, {}_pF_q(a;b;\hat{\mathcal{A}}_+\hat{\mathcal{A}}_-) \qquad (3.29)$$

where we have

$$\frac{d^2z}{\pi} = \frac{d\varphi}{2\pi} d(|z|^2) \qquad (3.30)$$

The integral over the angular variable will be

$$\int_0^{2\pi} \frac{d\varphi}{2\pi} {}_pF_q(a;b;z\hat{\mathcal{A}}_+) \, {}_pF_q(a;b;z^*\hat{\mathcal{A}}_-) = \sum_{n=0}^{\infty} \frac{(\hat{\mathcal{A}}_+\hat{\mathcal{A}}_-)^n}{[\rho_{p,q}(b/a \,|\, n)]^2} (|z|^2)^n =$$

$$= {}_{2p}F_{2q+1}(a,a;1,b,b;|z|^2 \hat{\mathcal{A}}_+\hat{\mathcal{A}}_-) \qquad (3.31)$$

so that the integral in real space, after the variable $|z|^2$ will be

$$\int_0^{R_c \le \infty} d(|z|^2) G_{p,q+1}^{q+1,0}\!\left(|z|^2 \left| \begin{array}{cc} / \; ; & a-1 \\ 0, \; b-1 \; ; & / \end{array} \right.\right) {}_{2p}F_{2q+1}(a,a;1,b,b;|z|^2 \hat{\mathcal{A}}_+\hat{\mathcal{A}}_-) =$$

$$= \frac{\prod_{j=1}^{q} \Gamma(b_j)}{\prod_{i=1}^{p} \Gamma(a_i)} \, {}_pF_q(a;b;\hat{\mathcal{A}}_+\hat{\mathcal{A}}_-) \qquad (3.32)$$

*Note*: The domain of integration with finite radius of convergence $R_c < \infty$, which involves Meijer's G functions, can be extended for $R_c = \infty$, if the step function of Heaviside is introduced:

$$H(R_c - |z|^2) = \begin{cases} 0, & |z|^2 > R_c \\ 1, & |z|^2 \le R_c \end{cases}. \qquad (3.33)$$

so we will have, for example:

$$\int_0^{R_c < \infty} d(|z|^2) G_{p,q}^{m,n}(|z|^2 \,|\, ...) F(|z|^2) = \int_0^{R_c = \infty} d(|z|^2) G_{p,q}^{m,n}(|z|^2 \,|\, ...) H(R_c - |z|^2) F(|z|^2) \qquad (3.34)$$

The above integrals are absolutely general, being valid for all GH-CSs. Let us specify it for the CSs of the one-dimensional harmonic oscillator (HO-1D). But first, let's recall the commutation relation between the canonical annihilation $\hat{\mathcal{A}}_- \to \hat{a}$ and creation $\hat{\mathcal{A}}_+ \to \hat{a}^+$ operators, as well as $p = q = 0$, $a = b$ and the particle number operator $\hat{\mathcal{N}}$:

$$\hat{a}|n> = \sqrt{n}\,|n-1>, \quad \hat{a}^+|n> = \sqrt{n+1}\,|n+1>, \quad [\hat{a},\hat{a}^+] = \hat{a}\hat{a}^+ - \hat{a}^+\hat{a} = 1,$$

$$<n|\hat{a}^+\hat{a}|n> = n, \quad \hat{\mathcal{N}} = \hat{a}^+\hat{a}, \quad \hat{\mathcal{N}}|n> = n|n>$$

(3.35)

Consequently, the hypergeometrical functions are

$$_0F_0(\,;\,;x) = \exp(x), \quad G_{0,1}^{1,0}\!\left(x \left|\begin{array}{c}/\;;\;/\\0,\;;\;/\end{array}\right.\right) = \exp(-x)$$

(3.36)

The integral in complex space then becomes [17]

$$\int \frac{d^2z}{\pi}\exp(-|z|^2 + z\hat{a}^+ + z^*\hat{a}) = \exp(\hat{a}^+\hat{a}), \quad \int \frac{d^2z}{\pi}\exp[-(z^*-\hat{a}_+)(z-\hat{a}_-)] = 1$$

(3.37)

On the other hand, the integral in real space of variable $|z|^2$, for HO-1D is

$$\int_0^\infty d(|z|^2)\,e^{-|z|^2}\,_0F_1(\,;1;|z|^2\,\hat{a}^+\hat{a}) = \int_0^\infty d(|z|^2)\,e^{-|z|^2}\,I_0(2|z|\sqrt{\hat{a}^+\hat{a}}) =$$

$$= 2\int_0^\infty d(|z|)\,e^{-|z|^2}\,|z|\,J_0(-2i|z|\sqrt{\hat{a}^+\hat{a}}) = e^{\hat{a}^+\hat{a}} = {}_0F_0(\,;\,;\hat{a}^+\hat{a})$$

(3.38)

where we have used the relation between the Bessel functions of the first kind with the modified Bessel functions of the first kind $J_\nu(x) = \dfrac{x^\nu}{(ix)^\nu}I_\nu(ix)$, as well as the integral of kind 6.631.4, page 706 of the Gradshteyn book [21].

Let us now consider another generalized hypergeometric function, where $C$ is a constant:

$$_rF_s(\pmb{c};\pmb{d};C|z|^2) = \sum_{n=0}^\infty \frac{\prod_{i=1}^r (c_i)_n}{\prod_{j=1}^s (d_j)_n}\frac{(C|z|^2)^n}{n!} \equiv \sum_{n=0}^\infty \frac{1}{\rho_{r,s}(d/c\,|\,n)}(C|z|^2)^n$$

(3.39)

and calculate the integral from real space (for deduction, see Appendix A 1)

$$\int_0^{R_c\leq\infty} d(|z|^2)\,G_{p,q+1}^{q+1,0}\!\left(|z|^2\left|\begin{array}{c}/\;;\;\pmb{a}-1\\0,\pmb{b}-1\;;\;/\end{array}\right.\right){}_rF_s(\pmb{c};\pmb{d};C|z|^2) =$$

$$= \frac{\prod_{j=1}^q \Gamma(b_j)}{\prod_{i=1}^p \Gamma(a_i)}\,{}_{q+r+1}F_{p+s}(1,\pmb{b},\pmb{c};\pmb{a},\pmb{d};C)$$

(3.40)

We previously emphasized that, within the DOOT formalism, the operators can be considered as simple classical numbers, so in the above equation we can take $C = \hat{\mathcal{A}}_+\hat{\mathcal{A}}_-$, $C = \vartheta_{\hat{\mathcal{A}}_+\hat{\mathcal{A}}_-}$ or $C = \vartheta_{|z|^2} \equiv |z|^2\,\dfrac{\partial}{\partial|z|^2}$. For the latter case, the integral above will be written as follows:

$$\int_0^{R_c\leq\infty} d(|z|^2)\, G_{p,q+1}^{q+1,0}\left(|z|^2 \,\Bigg|\, \begin{array}{c} / \;\; ; \;\; \boldsymbol{a-1} \\ 0,\, \boldsymbol{b-1}\; ;\;\; / \end{array}\right) {}_rF_s(\boldsymbol{c};\boldsymbol{d};\,\vartheta_{|z|^2}|z|^2) =$$

$$= \frac{\prod_{j=1}^{q}\Gamma(b_j)}{\prod_{i=1}^{p}\Gamma(a_i)}\, {}_{q+r+1}F_{p+s}(1,\boldsymbol{b},\boldsymbol{c};\,\boldsymbol{a},\boldsymbol{d};\,\vartheta_{|z|^2}) \tag{3.41}$$

Since generalized hypergeometric functions can be represented by more general or equivalent functions, by customizing the integer indices $p$, $q$, $r$, $s$, as well as the sets of real numbers $\boldsymbol{a},\boldsymbol{b},\boldsymbol{c},\boldsymbol{d}$, different integrals involving special or usual functions can be obtained.

For illustration, we will consider a single example, taking $p=1$, $q=0$, $r=1$, $s=1$ and $\boldsymbol{a}=a+1$, $\boldsymbol{b}=1$, $\boldsymbol{c}=\boldsymbol{d}=c$:

$$G_{1,1}^{1,0}\left(|z|^2 \,\Bigg|\, \begin{array}{c} / \;\; ; \;\; a \\ 0 \;\; ; \;\; / \end{array}\right) = \frac{1}{\Gamma(a)}(1-|z|^2)^{a-1}\,,\quad {}_1F_1(c;c;\,C|z|^2) = {}_0F_0(;\,;C|z|^2) = \exp(C|z|^2) \tag{3.42}$$

The corresponding integral is

$$\frac{1}{\Gamma(a)}\int_0^{R_c\leq\infty} d(|z|^2)(1-|z|^2)^{a-1}\exp(C|z|^2) =$$

$$= \frac{1}{\Gamma(a)}{}_2F_2(1,c;\,a+1,c;\,C|z|^2) = \frac{1}{\Gamma(a)}{}_1F_1(1;\,a+1;\,C|z|^2) \tag{3.43}$$

Thus, we obtained an integral representation of the Kummer confluent function ${}_1F_1$ on the real axis [22]:

$${}_1F_1(1;\,a+1;\,C|z|^2) = \int_0^1 d(|z|^2)\, e^{C|z|^2}(1-|z|^2)^{a-1} \tag{3.44}$$

If we choose $C = \vartheta_{|z|^2}$, and take into account that

$$e^{\vartheta_{|z|^2}|z|^2} = \sum_{m=0}^{\infty}\frac{1}{m!}(\vartheta_{|z|^2}|z|^2)^m = \sum_{m=0}^{\infty}\frac{1}{m!}(|z|^2)^m = e^{|z|^2} \tag{3.45}$$

we will get an interesting relationship related to the action of the theta operator:

$${}_1F_1(1;\,a+1;\,\vartheta_{|z|^2}|z|^2) = {}_1F_1(1;\,a+1;\,|z|^2) \tag{3.46}$$

Also, Eq. (3.40) leads to the definition of the Laplace transform a functiilor hypergeometrice:

$$\mathcal{L}\{{}_rF_s\}(S) = \int_0^{\infty} dt\, e^{-St}\, {}_rF_s(\boldsymbol{c};\boldsymbol{d};\,t) \tag{3.47}$$

Since Meijer's G-function must be reduced to an exponential, we will first change the variable $C|z|^2 \equiv t$, $C^{-1} \equiv S$, so that the integral (3.40) becomes:

$$\int_0^{R_c \leq \infty} dt\, G_{p,q+1}^{q+1,0}\left(St \left| \begin{array}{c} /\ ;\ \pmb{a}-\pmb{1} \\ 0,\ \pmb{b}-\pmb{1}\ ;\ / \end{array}\right.\right) {}_rF_s(\pmb{c};\pmb{d};t) =$$

$$= \frac{1}{S} \frac{\prod_{j=1}^{q}\Gamma(b_j)}{\prod_{i=1}^{p}\Gamma(a_i)} {}_{q+r+1}F_{p+s}\left(1,\pmb{b},\pmb{c};\pmb{a},\pmb{d};S^{-1}\right) \quad (3.48)$$

Then we will choose $p = 0$, $q = 0$, $\pmb{a} = \pmb{1}$, $\pmb{b} = \pmb{1}$ and obtain

$$\mathcal{L}\{{}_rF_s\}(S) = \int_0^\infty dt\, e^{-St}\, {}_rF_s(\pmb{c};\pmb{d};t) = = \frac{1}{S^{r+1}} {}_{r+1}F_s\left(1,\pmb{c};\pmb{d};S^{-1}\right) \quad (3.49)$$

Seen from another angle, Eq. (3.40) actually represents the integral representation of the hypergeometric function that appears on the right side.

### 4. The role of the operator $\vartheta_x$

To highlight the role of the theta operator $\vartheta_x = x\dfrac{\partial}{\partial x}$ and implicitly to simplify the writing of the formulas, we will use the following correspondence:

$$n\left(\hat{\mathcal{A}}_+\hat{\mathcal{A}}_-\right)^n \Leftrightarrow \vartheta_{\hat{\mathcal{A}}_+\hat{\mathcal{A}}_-}\left(\hat{\mathcal{A}}_+\hat{\mathcal{A}}_-\right)^n \equiv \hat{\mathcal{A}}_+\hat{\mathcal{A}}_-\frac{\partial}{\partial \hat{\mathcal{A}}_+\hat{\mathcal{A}}_-}\left(\hat{\mathcal{A}}_+\hat{\mathcal{A}}_-\right)^n \quad (4.01)$$

which means that, from the operational point of view, if a function depends on the operational "variable" $\hat{\mathcal{A}}_+\hat{\mathcal{A}}_-$, the operator $\vartheta_{\hat{\mathcal{A}}_+\hat{\mathcal{A}}_-}$ can be applied to it, respecting the rules indicated in Section 2.

In this sense, the expressions of the energy eigenvalues, and the structure functions, they will be subject to the following correspondence:

$$e(n) = n\frac{\prod_{j=1}^{q}(b_j + n - 1)}{\prod_{i=1}^{p}(a_i + n - 1)} \quad \Leftrightarrow \quad e(\vartheta_{\hat{\mathcal{A}}_+\hat{\mathcal{A}}_-}) = \vartheta_{\hat{\mathcal{A}}_+\hat{\mathcal{A}}_-}\frac{\prod_{j=1}^{q}(b_j + \vartheta_{\hat{\mathcal{A}}_+\hat{\mathcal{A}}_-} - 1)}{\prod_{i=1}^{p}(a_i + \vartheta_{\hat{\mathcal{A}}_+\hat{\mathcal{A}}_-} - 1)} \quad (4.02)$$

$$\rho_{p,q}(b/a\,|\,n) = n!\frac{\prod_{j=1}^{q}(b_j)_n}{\prod_{i=1}^{p}(a_i)_n} \quad \Leftrightarrow \quad \rho_{p,q}(b/a\,|\,\vartheta_{\hat{\mathcal{A}}_+\hat{\mathcal{A}}_-}) = (\vartheta_{\hat{\mathcal{A}}_+\hat{\mathcal{A}}_-})!\frac{\prod_{j=1}^{q}(b_j)_{\vartheta_{\hat{\mathcal{A}}_+\hat{\mathcal{A}}_-}}}{\prod_{i=1}^{p}(a_i)_{\vartheta_{\hat{\mathcal{A}}_+\hat{\mathcal{A}}_-}}} \quad (4.03)$$

Now, let's apply the operator $\vartheta_{\hat{\mathcal{A}}_+\hat{\mathcal{A}}_-}$ on the generalized hypergeometric function

$$\vartheta_{\hat{\mathcal{A}}_+\hat{\mathcal{A}}_-}\, _pF_q(a;b;\hat{\mathcal{A}}_+\hat{\mathcal{A}}_-) = \sum_{n=0}^{\infty} n\, \frac{\prod_{i=1}^{p}(a_i)_n}{\prod_{j=1}^{q}(b_j)_n} \frac{(\hat{\mathcal{A}}_+\hat{\mathcal{A}}_-)^n}{n!} = \sum_{n=0}^{\infty} \frac{\prod_{i=1}^{p}(a_i+n-1)\prod_{i=1}^{p}(a_i)_{n-1}}{\prod_{j=1}^{q}(b_j+n-1)\prod_{j=1}^{q}(b_j)_{n-1}} \frac{(\hat{\mathcal{A}}_+\hat{\mathcal{A}}_-)^n}{(n-1)!} \quad (4.04)$$

where a property of Pohhammer's symbol was used.

Using the relationship

$$n^m\,(\hat{\mathcal{A}}_+\hat{\mathcal{A}}_-)^n = (\vartheta_{\hat{\mathcal{A}}_+\hat{\mathcal{A}}_-})^m (\hat{\mathcal{A}}_+\hat{\mathcal{A}}_-)^n \quad (4.05)$$

we will be able to remove the first fraction under the sum sign. In addition, changing the summation index from the remaining sum, $m=n-1$ and giving up the non-physical term $m=-1$, we obtain

$$\vartheta_{\hat{\mathcal{A}}_+\hat{\mathcal{A}}_-}\, _pF_q(a;b;\hat{\mathcal{A}}_+\hat{\mathcal{A}}_-) = (\hat{\mathcal{A}}_+\hat{\mathcal{A}}_-)\frac{\prod_{i=1}^{p}(a_i+\vartheta_{\hat{\mathcal{A}}_+\hat{\mathcal{A}}_-}-1)}{\prod_{j=1}^{q}(b_j+\vartheta_{\hat{\mathcal{A}}_+\hat{\mathcal{A}}_-}-1)}\sum_{m=0}^{\infty}\frac{\prod_{i=1}^{p}(a_i)_m}{\prod_{j=1}^{q}(b_j)_m}\frac{(\hat{\mathcal{A}}_+\hat{\mathcal{A}}_-)^m}{m!} =$$

$$= (\hat{\mathcal{A}}_+\hat{\mathcal{A}}_-)\frac{\prod_{i=1}^{p}(a_i+\vartheta_{\hat{\mathcal{A}}_+\hat{\mathcal{A}}_-}-1)}{\prod_{j=1}^{q}(b_j+\vartheta_{\hat{\mathcal{A}}_+\hat{\mathcal{A}}_-}-1)}\, _pF_q(a;b;\hat{\mathcal{A}}_+\hat{\mathcal{A}}_-) \quad (4.06)$$

Similarly, it can be shown that the relation below is also valid:

$$\vartheta_{\hat{\mathcal{A}}_+\hat{\mathcal{A}}_-} \frac{\prod_{j=1}^{q}(b_j+\vartheta_{\hat{\mathcal{A}}_+\hat{\mathcal{A}}_-}-1)}{\prod_{i=1}^{p}(a_i+\vartheta_{\hat{\mathcal{A}}_+\hat{\mathcal{A}}_-}-1)}\, _pF_q(a;b;\hat{\mathcal{A}}_+\hat{\mathcal{A}}_-) = (\hat{\mathcal{A}}_+\hat{\mathcal{A}}_-)\, _pF_q(a;b;\hat{\mathcal{A}}_+\hat{\mathcal{A}}_-) \quad (4.07)$$

If we use the notations

$$\mathcal{A}(\vartheta_{\hat{\mathcal{A}}_+\hat{\mathcal{A}}_-}) \equiv \prod_{i=1}^{p}(a_i+\vartheta_{\hat{\mathcal{A}}_+\hat{\mathcal{A}}_-}-1) \quad , \quad \mathcal{B}(\vartheta_{\hat{\mathcal{A}}_+\hat{\mathcal{A}}_-}) \equiv \prod_{j=1}^{q}(b_j+\vartheta_{\hat{\mathcal{A}}_+\hat{\mathcal{A}}_-}-1) \quad (4.07)$$

and the differential equation for generalized hypergeometric functions [23]

$$\left[\vartheta_{\hat{\mathcal{A}}_+\hat{\mathcal{A}}_-}\mathcal{B}(\vartheta_{\hat{\mathcal{A}}_+\hat{\mathcal{A}}_-}) - \hat{\mathcal{A}}_+\hat{\mathcal{A}}_-\mathcal{A}(\vartheta_{\hat{\mathcal{A}}_+\hat{\mathcal{A}}_-})\right]\, _pF_q(a;b;\hat{\mathcal{A}}_+\hat{\mathcal{A}}_-) = 0 \quad (4.08)$$

by formal division by $\mathcal{B}(\vartheta_{\hat{\mathcal{A}}_+\hat{\mathcal{A}}_-})$, we can verify that the above relation is correct.

$$\left[\vartheta_{\hat{\mathcal{A}}_+\hat{\mathcal{A}}_-} - \hat{\mathcal{A}}_+\hat{\mathcal{A}}_-\frac{\mathcal{A}(\vartheta_{\hat{\mathcal{A}}_+\hat{\mathcal{A}}_-})}{\mathcal{B}(\vartheta_{\hat{\mathcal{A}}_+\hat{\mathcal{A}}_-})}\right]\, _pF_q(a;b;\hat{\mathcal{A}}_+\hat{\mathcal{A}}_-) = 0 \quad (4.09)$$

Let us now proceed to the examination of *mixed states*. We will refer to the case of the thermal states of a quantum system in canonical equilibrium with the ambient environment ("bath") at temperature $T = (k_B \beta)^{-1}$, where $k_B$ is the Boltzman's constant, and $\beta$ the less dimensional parameter of the temperature). Mixed states are described by the density operator

$$\hat{\rho} = \frac{1}{Z(\beta)} \exp(-\beta \hat{\mathcal{H}}) = \frac{1}{Z(\beta)} \sum_n \exp[-\beta \hbar \omega e(n)] | n \rangle \langle n | \qquad (4.10)$$

with the partition function

$$Z(\beta) = \sum_n \exp[-\beta \hbar \omega e(n)] \qquad (4.11)$$

Using the properties of the $\hat{\mathcal{A}}_+$ and $\hat{\mathcal{A}}_-$ operators, as well as the DOOT rules, the density operator can be written as a function which depend only of the product $\hat{\mathcal{A}}_+ \hat{\mathcal{A}}_-$:

$$\hat{\rho} = \frac{1}{Z(\beta)} \sum_n \exp[-\beta \hbar \omega e(n)] \frac{(\hat{\mathcal{A}}_+)^n | 0 \rangle \langle 0 | (\hat{\mathcal{A}}_-)^n}{\rho_{p,q}(b/a | n)} =$$
$$= \frac{1}{Z(\beta)} \frac{1}{{}_pF_q(a; b; \hat{\mathcal{A}}_+ \hat{\mathcal{A}}_-)} \sum_n \exp[-\beta \hbar \omega e(n)] \frac{(\hat{\mathcal{A}}_+ \hat{\mathcal{A}}_-)^n}{\rho_{p,q}(b/a | n)} \qquad (4.12)$$

which is equivalent to

$$\hat{\rho} = \frac{1}{Z(\beta)} \frac{1}{{}_pF_q(a; b; \hat{\mathcal{A}}_+ \hat{\mathcal{A}}_-)} \exp[-\beta \hbar \omega \, e(\vartheta_{\hat{\mathcal{A}}_+ \hat{\mathcal{A}}_-})] \, {}_pF_q(a; b; \hat{\mathcal{A}}_+ \hat{\mathcal{A}}_-) \qquad (4.13)$$

The coherent states in the Barut-Girardello manner (BG-CSs) are eigenvectors of the annihilation operator $\hat{\mathcal{A}}_-$, having the eigenvalue $z$ [24]. The corresponding definition, together with the Hermitically conjugate one, is defined as follows:

$$\hat{\mathcal{A}}_- | z \rangle = z | z \rangle \quad , \quad \langle z^* | \hat{\mathcal{A}}_+ = z^* \langle z^* | \qquad (4.13)$$

so that the scalar product of the two vectors is

$$\langle z^* | \hat{\mathcal{A}}_+ \hat{\mathcal{A}}_- | z' \rangle = z^* z' \langle z^* | z' \rangle = z^* z' \frac{{}_pF_q(a; b; z^* z')}{\sqrt{{}_pF_q(a; b; |z|^2)} \sqrt{{}_pF_q(a; b; |z'|^2)}} \qquad (4.14)$$

Assuming that, in the general case, the eigenvalue equation below is valid

$$\hat{\mathcal{F}}(\hat{\mathcal{A}}_+ \hat{\mathcal{A}}_-) | n \rangle = f(n) | n \rangle \quad , \quad \hat{\mathcal{F}}(\hat{\mathcal{A}}_+ \hat{\mathcal{A}}_-) = \sum_n f(n) | n \rangle \langle n | \qquad (4.15)$$

for the matrix elements in the GH CSs representation we will obtain:

$$\langle z^* | \hat{\mathcal{F}}(\hat{\mathcal{A}}_+ \hat{\mathcal{A}}_-) | z' \rangle = \sum_n f(n) \frac{(z^* z')^n}{\rho_{p,q}(b/a | n)} = \sum_n f(\vartheta_{z^*}) \frac{(z^* z')^n}{\rho_{p,q}(b/a | n)} =$$
$$= f(\vartheta_{z^*}) \sum_n \frac{(z^* z')^n}{\rho_{p,q}(b/a | n)} = \hat{\mathcal{F}}(\vartheta_{z^*}) \, {}_pF_q(a; b; z^* z') \qquad (4.16)$$

in which it is observed that each main quantum number $n$ in the energy eigenvalue expression has been replaced by the operator $\vartheta_{z^*}$, i.e. $n \Leftrightarrow \vartheta_{z^*} = z^* \dfrac{\partial}{\partial z^*} = z^* z' \dfrac{\partial}{\partial z^* z'} = \vartheta_{z^* z'}$. Consequently, the obtained operational expression can be removed from under the sum sign.

Moreover, in the BG-CSs representation the density operator is represented by the following density matrix

$$<z^*|\hat{\rho}|z'> = \frac{1}{Z(\beta)} <z^*| \frac{1}{{}_pF_q(a; b; \hat{\mathcal{A}}_+\hat{\mathcal{A}}_-)} \exp\left[-\beta\hbar\omega\, e(\vartheta_{\hat{\mathcal{A}}_+\hat{\mathcal{A}}_-})\right] {}_pF_q(a; b; \hat{\mathcal{A}}_+\hat{\mathcal{A}}_-)|z'> \quad (4.17)$$

or, according to Eq. (4.10) and the expression above, the density matrix can also be written

$$<z^*|\hat{\rho}(\hat{\mathcal{A}}_+\hat{\mathcal{A}}_-)|z'> = \hat{\rho}(\vartheta_{z^*})\, {}_pF_q(a; b; z^*z') =$$
$$= \frac{1}{Z(\beta)} \frac{1}{{}_pF_q(a; b; z^*z')} \exp\left[-\beta\hbar\omega\, e(\vartheta_{z^*})\right] {}_pF_q(a; b; z^*z') \quad (4.18)$$

These expressions are too complicated, in which each product $<z^*|\hat{\mathcal{A}}_+\hat{\mathcal{A}}_-|z'>$ should be replaced with the right member of Eq. (4.14). Therefore, the procedure is not convenient from a practical point of view. It is considerably simplified for the diagonal elements, where each product $<z^*|\hat{\mathcal{A}}_+\hat{\mathcal{A}}_-|z>$ will be replaced by $|z|^2$. We implicitly arrive at the expression of Husimi's distribution function [25 Husimi]:

$$Q^{(H)}(|z|^2) \equiv <z^*|\hat{\rho}|z> = \frac{1}{Z(\beta)} \frac{1}{{}_pF_q(a; b; |z|^2)} \exp\left[-\beta\hbar\omega\, e(\vartheta_{|z|^2})\right] {}_pF_q(a; b; |z|^2) \quad (4.19)$$

The density operator can also be expressed as a weighted superposition of projectors of coherent states:

$$\hat{\rho} = \frac{1}{Z(\beta)} \int d\mu(z) P(|z|^2; \beta)|z><z^*| \quad (4.20)$$

where $P(|z|^2; \beta)$ is a quasi-distribution function, because on certain intervals it can also take negative values.

The corresponding expression that uses operators is the one below, in which we will have to find the function $P(|z|^2; \beta)$:

$$\hat{\rho} = \frac{1}{Z(\beta)} \frac{\prod_{i=1}^{p} \Gamma(a_i)}{\prod_{j=1}^{q} \Gamma(b_j)} \frac{1}{{}_pF_q(a; b; \hat{\mathcal{A}}_+\hat{\mathcal{A}}_-)} \int_0^{R_c} d(|z|^2)\, G_{p,q+1}^{q+1,0}\!\left(|z|^2 \left| \begin{array}{c} /\ ;\ a-1 \\ 0,\ b-1\ ;\ / \end{array}\right.\right) P(|z|^2; \beta) \times \quad (4.21)$$
$$\times \int_0^{2\pi} \frac{d\varphi}{2\pi}\, {}_pF_q(a; b; z\hat{\mathcal{A}}_+)\, {}_pF_q(a; b; z^*\hat{\mathcal{A}}_-)$$

After calculating the angular integral and changing the function

$$\tilde{P}(|z|^2 ; \beta) \equiv G_{p,q+1}^{q+1,0}\left(|z|^2 \left| \begin{array}{c} /\ ;\ \ \ \ a-1 \\ 0,\ b-1\ ;\ \ /\end{array}\right.\right) P(|z|^2 ; \beta) \quad (4.22)$$

we will get

$$\hat{\rho} = \frac{1}{Z(\beta)} \frac{\prod_{i=1}^{p}\Gamma(a_i)}{\prod_{j=1}^{q}\Gamma(b_j)} \frac{1}{{}_pF_q(a;b;\hat{\mathcal{A}}_+\hat{\mathcal{A}}_-)} \sum_{n=0}^{\infty} \frac{(\hat{\mathcal{A}}_+\hat{\mathcal{A}}_-)^n}{[\rho_{p,q}(b/a|n)]^2} \int_0^{R_c} d(|z|^2) \tilde{P}(|z|^2 ; \beta)(|z|^2)^n \quad (4.23)$$

Let's equal the factors on the right side of equations (4.12) and (4.20). We will have:

$$\sum_n \exp[-\beta\hbar\omega e(n)] \frac{(\hat{\mathcal{A}}_+\hat{\mathcal{A}}_-)^n}{\rho_{p,q}(b/a|n)} = \frac{\prod_{i=1}^{p}\Gamma(a_i)}{\prod_{j=1}^{q}\Gamma(b_j)} \sum_{n=0}^{\infty} \frac{(\hat{\mathcal{A}}_+\hat{\mathcal{A}}_-)^n}{[\rho_{p,q}(b/a|n)]^2} \int_0^{R_c} d(|z|^2) \tilde{P}(|z|^2 ; \beta)(|z|^2)^n \quad (4.24)$$

It is observed that in order to have equality, the integral must be

$$\int_0^{R_c} d(|z|^2) \tilde{P}(|z|^2 ; \beta)(|z|^2)^n = \frac{\prod_{j=1}^{q}\Gamma(b_j)}{\prod_{i=1}^{p}\Gamma(a_i)} \rho_{p,q}(b/a|n) \exp[-\beta\hbar\omega e(n)] \quad (4.25)$$

The concrete expression of the quasi-distribution function $P(|z|^2 ; \beta)$ can only be found after the concrete explanation of the expression of the energy eigenvalues $e(n)$. Below we will exemplify this statement for the case of *systems with a linear energy spectrum*.

Let us consider, therefore, a quantum system characterized by a linear energy spectrum, i.e. one in which the eigenvalues of the system's energy depend linearly on the main quantum number $n$. For this case $p = q = 1$ and also $a = 1$; $b = e_0 + 1$, $\rho_{1,1}(e_0+1|n) = (e_0+1)_n$.

To distinguish it from the general case, we will note the characteristic quantities for the linear spectrum with the superscript $(L)$.

Following the standard procedure, it is obtained that GH- CSs of the Barut-Girardello type have the following expression:

$$|z>^{(L)} = \frac{1}{\sqrt{{}_1F_1(1; e_0+1; |z|^2)}} \sum_{n=0}^{\infty} \frac{z^n}{\sqrt{(e_0+1)_n}} |n> =$$
$$= \frac{1}{\sqrt{{}_1F_1(1; e_0+1; |z|^2)}} {}_1F_1(1; e_0+1; z\hat{\mathcal{A}}_+)|0> \quad (4.26)$$

and similar their complex conjugate. For this case, the generalized hypergeometric function is

$$_1F_1(1; e_0+1; z^*z') = \sum_{n=0}^{\infty} \frac{(1)_n}{(e_0+1)_n} \frac{(z^*z')^n}{n!} \tag{4.27}$$

The projector on the GH CSs is

$$|z>^{(L)\ (L)}<z^*| = \frac{1}{_1F_1(1; e_0+1; |z|^2)} \frac{_1F_1(1; e_0+1; z\hat{A}_+)\,_1F_1(1; e_0+1; z^*\hat{A}_-)}{_1F_1(1; e_0+1; \hat{A}_+\hat{A}_-)} \tag{4.28}$$

while the integration measure has the expression

$$d\mu^{(L)}(z) = \frac{1}{\Gamma(e_0+1)} \frac{d\varphi}{2\pi} d(|z|^2) G_{1,2}^{2,0}\left(|z|^2 \left|\begin{array}{c} /\ ;\ \ 0 \\ 0,\ e_0\ ;\ / \end{array}\right.\right) {}_1F_1(1; e_0+1; |z|^2) =$$

$$= \frac{1}{\Gamma(e_0+1)} \frac{d\varphi}{2\pi} d(|z|^2) G_{0,1}^{1,0}(|z|^2\ |\ e_0)\,_1F_1(1; e_0+1; |z|^2) = \tag{4.29}$$

$$= \frac{1}{\Gamma(e_0+1)} \frac{d\varphi}{2\pi} d(|z|^2) e^{-|z|^2} (|z|^2)^{e_0}\,_1F_1(1; e_0+1; |z|^2)$$

The integral over the angular variable will be

$$\int_0^{2\pi} \frac{d\varphi}{2\pi}\,_1F_1(1; e_0+1; z\hat{A}_+)\,_1F_1(1; e_0+1; z^*\hat{A}_-) = {}_2F_3(1,1; 1, e_0+1, e_0+1; |z|^2\,\hat{A}_+\hat{A}_-) =$$

$$= {}_1F_2(1; e_0+1, e_0+1; |z|^2\,\hat{A}_+\hat{A}_-) = \sum_{n=0}^{\infty} \frac{(1)_n}{[(e_0+1)_n]^2} \frac{(|z|^2\,\hat{A}_+\hat{A}_-)^n}{n!} \tag{4.30}$$

From the decomposition relation of the unity operator and after performing the angular integral, we get the following important integral, according to Eq. (3.32).

$$\int_0^{R_c \leq \infty} d(|z|^2) G_{1,2}^{2,0}\left(|z|^2 \left|\begin{array}{c} /\ ;\ \ 0 \\ 0,\ e_0\ ;\ / \end{array}\right.\right) {}_2F_3(1,1; 1, e_0+1, e_0+1; |z|^2\,\hat{A}_+\hat{A}_-) =$$

$$= \int_0^{R_c \leq \infty} d(|z|^2) G_{0,1}^{1,0}(|z|^2\ |\ e_0)\,_1F_2(1; e_0+1, e_0+1; |z|^2\,\hat{A}_+\hat{A}_-) =$$

$$= \sum_{n=0}^{\infty} \frac{(1)_n}{[(e_0+1)_n]^2} \frac{(\hat{A}_+\hat{A}_-)^n}{n!} \int_0^{R_c \leq \infty} d(|z|^2) e^{-|z|^2} (|z|^2)^{n+e_0} = \sum_{n=0}^{\infty} \frac{(1)_n}{[(e_0+1)_n]^2} \frac{(\hat{A}_+\hat{A}_-)^n}{n!} \Gamma(e_0+1+n) = \tag{4.31}$$

$$= \Gamma(e_0+1) \sum_{n=0}^{\infty} \frac{(1)_n}{(e_0+1)_n} \frac{(\hat{A}_+\hat{A}_-)^n}{n!} = \Gamma(e_0+1)\,_1F_1(1; e_0+1; \hat{A}_+\hat{A}_-)$$

So, generally we can write the integral

$$\int \frac{d^2z}{\pi} G_{1,2}^{2,0}\left(|z|^2 \left| \begin{array}{c} / \; ; \; 0 \\ 0, \; e_0 \; ; \; / \end{array}\right.\right) {}_1F_1(1; e_0+1; z\hat{\mathcal{A}}_+) {}_1F_1(1; e_0+1; z^*\hat{\mathcal{A}}_-) = $$
$$= \Gamma(e_0+1) {}_1F_1(1; e_0+1; \hat{\mathcal{A}}_+\hat{\mathcal{A}}_-) \quad (4.32)$$

The mixed states of such a system are characterized by the density operator

$$\hat{\rho}^{(L)} = \frac{1}{Z^{(L)}(\beta)} e^{-\beta\hbar\omega e_0} \sum_{n=0}^{\infty} \left(e^{-\beta\hbar\omega}\right)^n |n\rangle\langle n| \quad (4.33)$$

where the partition function is

$$Z^{(L)}(\beta) = e^{-\beta\hbar\omega e_0} \sum_{n=0}^{\infty} \left(e^{-\beta\hbar\omega}\right)^n = \frac{e^{-\beta\hbar\omega e_0}}{1-e^{-\beta\hbar\omega}} \quad (4.34)$$

As a function of the product of $\hat{\mathcal{A}}_+\hat{\mathcal{A}}_-$ operators, using the DOOT rules, the density operator is written as

$$\hat{\rho}^{(L)} = \frac{1}{Z^{(L)}(\beta)} e^{-\beta\hbar\omega e_0} \frac{{}_1F_1(1; e_0+1; e^{-\beta\hbar\omega}\hat{\mathcal{A}}_+\hat{\mathcal{A}}_-)}{{}_1F_1(1; e_0+1; \hat{\mathcal{A}}_+\hat{\mathcal{A}}_-)} \quad (4.35)$$

For systems with a linear energy spectrum, the correspondence $n \Leftrightarrow \vartheta_{\hat{\mathcal{A}}_+\hat{\mathcal{A}}_-}$ will have as a consequence the relationship

$$e(n) = n + e_0 \quad \Leftrightarrow \quad e(\vartheta_{\hat{\mathcal{A}}_+\hat{\mathcal{A}}_-}) = \vartheta_{\hat{\mathcal{A}}_+\hat{\mathcal{A}}_-} + e_0 = \hat{\mathcal{A}}_+\hat{\mathcal{A}}_- \frac{\partial}{\partial \hat{\mathcal{A}}_+\hat{\mathcal{A}}_-} + e_0 \quad (4.36)$$

It means that the density operator can be written in another equivalent form

$$\hat{\rho}^{(L)} = \frac{1}{Z(\beta)} \frac{1}{{}_1F_1(1; e_0+1; \hat{\mathcal{A}}_+\hat{\mathcal{A}}_-)} \exp\left[-\beta\hbar\omega \; e(\vartheta_{\hat{\mathcal{A}}_+\hat{\mathcal{A}}_-})\right] {}_1F_1(1; e_0+1; \hat{\mathcal{A}}_+\hat{\mathcal{A}}_-) \quad (4.37)$$

The matrix elements of the density operator in the GH-CSs representation have the expression

$${}^{(L)}\langle z^*|\hat{\rho}^{(L)}|z'\rangle^{(L)} = \frac{1}{Z^{(L)}(\beta)} e^{-\beta\hbar\omega e_0} \frac{{}_1F_1(1; e_0+1; e^{-\beta\hbar\omega} z^*z')}{{}_1F_1(1; e_0+1; z^*z')} \quad (4.38)$$

Consequently, the diagonal elements, that is, Husimi's distribution function, is

$$Q_H^{(L)}(|z|^2) \equiv {}^{(L)}\langle z^*|\hat{\rho}^{(L)}|z\rangle^{(L)} = \frac{1}{Z^{(L)}(\beta)} e^{-\beta\hbar\omega e_0} \frac{{}_1F_1(1; e_0+1; e^{-\beta\hbar\omega} |z|^2)}{{}_1F_1(1; e_0+1; |z|^2)} \quad (4.39)$$

To evaluate the ratio of the above two generalized hypergeometric functions (also known as Kummer confluent hypergeometric functions) we will use the relation that expresses the product of two hypergeometric functions, whose argument differs only by a constant [26].

$$_pF_q(\pmb{a};\pmb{b};x)\,_rF_s(\pmb{c};\pmb{d};gx)=$$

$$=\sum_{m=0}^{\infty}\frac{\prod_{i=1}^{p}(a_i)_m}{\prod_{j=1}^{q}(b_j)_m}\,_{q+r+1}F_{p+s}\left(-m,1-m-\pmb{b},\pmb{c};1-m-\pmb{a};\pmb{d};(-1)^{p+q+1}g\right)\frac{x^m}{m!} \quad (4.40)$$

Taking into account the relationships

$$_2F_1(a,b;a;x)=\,_1F_0(b;\,;x)=\frac{1}{(1-x)^b}\quad,\quad _0F_0(\,;\,;\varsigma)=e^x \quad (4.41)$$

let's calculate the product of the functions below. We will have (see, Appendix A-2):

$$_1F_1(1;e_0+1;|z|^2)\,_0F_0(\,;\,;(e^{-\beta\hbar\omega}-1)|z|^2)=e^{-\beta\hbar\omega e_0}\,_1F_1(1;e_0+1;e^{-\beta\hbar\omega}z|^2) \quad (4.42)$$

Then the Husimi's distribution function becomes finally

$$Q_H^{(L)}(|z|^2)=\frac{1}{Z^{(L)}(\beta)}\exp\left[(e^{-\beta\hbar\omega}-1)\right] \quad (4.43)$$

Often, the expression of the Bose-Einstein distribution function is used in the literature

$$\bar{n}=\frac{1}{e^{\beta\hbar\omega}-1} \quad (4.44)$$

so that Husimi's distribution function will be written as

$$Q_H^{(L)}(|z|^2)=\frac{1}{\bar{n}+1}\left(\frac{\bar{n}+1}{\bar{n}}\right)^{e_0}\exp\left(-\frac{1}{\bar{n}+1}|z|^2\right) \quad (4.45)$$

which means that the behavior of the Husimi's distribution resembles that of a Gaussian distribution.

Let's now calculate the quasi-distribution function $P^{(L)}(|z|^2;\beta)$. After the standard change of the exponent $n=s-1$, equation (4.26) will now be written as

$$\int_0^{R_c\leq\infty}d(|z|^2)\,\tilde{P}^{(L)}(|z|^2;\beta)(|z|^2)^{s-1}=e^{-\beta\hbar\omega e_0}\,e^{\beta\hbar\omega}\frac{1}{(e^{\beta\hbar\omega})^s}\Gamma(e_0+s) \quad (4.46)$$

The solution of this moment equation is [15]

$$\tilde{P}^{(L)}(|z|^2;\beta)=e^{-\beta\hbar\omega e_0}\,e^{\beta\hbar\omega}G_{1,2}^{2,0}\left(e^{\beta\hbar\omega}|z|^2\,\middle|\,\begin{array}{cc}/\,;&0\\0,\,e_0\,;&/\end{array}\right)$$

$$=e^{-\beta\hbar\omega e_0}\,e^{\beta\hbar\omega}\,e^{-e^{\beta\hbar\omega}|z|^2}\left(e^{\beta\hbar\omega}|z|^2\right)^{e_0} \quad (4.47)$$

Finally, the quasi-distribution function $P^{(L)}(|z|^2;\beta)$ is

$$P^{(L)}\left(|z|^2 \ ; \ \beta\right) = \frac{1}{Z^{(L)}(\beta)} e^{-\beta\hbar\omega e_0} e^{\beta\hbar\omega} \frac{G_{1,2}^{2,0}\left(e^{\beta\hbar\omega} |z|^2 \ \bigg| \ \begin{array}{c} / \ ; \ 0 \\ 0, \ e_0 \ ; \ / \end{array}\right)}{G_{1,2}^{2,0}\left(|z|^2 \ \bigg| \ \begin{array}{c} / \ ; \ 0 \\ 0, \ e_0 \ ; \ / \end{array}\right)} =$$

$$= \left(e^{\beta\hbar\omega} - 1\right) \exp\left[-\left(e^{\beta\hbar\omega} - 1\right)|z|^2\right] = \frac{1}{\bar{n}} \exp\left(-\frac{1}{\bar{n}}|z|^2\right)$$

(4.48)

Consequently, the integral representation of the density operator (often called the diagonal representation) will be written as a weighted superposition of projectors of coherent states:

$$\hat{\rho}^{(L)} = \frac{1}{Z^{(L)}(\beta)} \int d\mu^{(L)}(z) \exp\left(-\frac{1}{\bar{n}}|z|^2\right) |z>^{(L)} \ {}^{(L)}<z^*| =$$

$$= \frac{1}{\Gamma(e_0+1)} \frac{1}{\bar{n}(\bar{n}+1)} \left(\frac{\bar{n}+1}{\bar{n}}\right)^{e_0} \frac{1}{{}_1F_1\left(1; \ e_0+1; \hat{\mathcal{A}}_+\hat{\mathcal{A}}_-\right)} \times$$

$$\times \int \frac{d^2z}{\pi} \exp\left(-\frac{\bar{n}+1}{\bar{n}}|z|^2\right) \left(|z|^2\right)^{e_0} {}_1F_1\left(1; \ e_0+1; z\hat{\mathcal{A}}_+\right) {}_1F_1\left(1; \ e_0+1; z^*\hat{\mathcal{A}}_-\right)$$

(4.49)

### 5. Concluding remarks

In the paper we examined the role and contribution of the theta operator $\vartheta_x \equiv x \frac{\partial}{\partial x}$ within the formalism of the most general coherent states - hypergeometric coherent states. Since any coherent state can be expanded in an orthogonal basis (in particular, the basis of Fock vectors), in which the expansion coefficients are proportional to the powers of the argument of the coherent state, the action of the theta operator on the expansion argument is particularly useful, that is: $\vartheta_x x^n = n \ x^n$. This allows to establish a correspondence between the main quantum number $n$ and the theta operator, $n \Leftrightarrow \vartheta_x$. As a consequence, the expression in which the theta operator appears can be removed from under the sum sign and thus simpler expressions can be obtained. In the paper, the DOOT technique (Diagonal Ordering Operation Technique) correlated with the $\hat{\mathcal{A}}_+$ creation and $\hat{\mathcal{A}}_-$ annihilation operators. With these, the theta operator was built in the form $\vartheta_{\hat{\mathcal{A}}_+\hat{\mathcal{A}}_-} \equiv \frac{\partial}{\partial \hat{\mathcal{A}}_+\hat{\mathcal{A}}_-}$ which acted on the functions depending on the ordered product of operators $\hat{\mathcal{A}}_+\hat{\mathcal{A}}_-$, using at the same time the DOOT rules. A series of new results were obtained and some already known ones were found / confirmed (the integral representations, as well as the Laplace transform of hypergeometric functions). We demonstrated that in the generalized hypergeometric coherent states representation each main quantum number $n$ in the expression of the energy eigenvalues is replaced by the theta operators $\vartheta_{z^*}$, $\vartheta_{|z|^2}$ or $\vartheta_{\hat{\mathcal{A}}_+\hat{\mathcal{A}}_-}$, as the case. The idea comes from the property of the operator's algebra in the complex Segal-Bargmann space, in which the algebra of ladder operators $\hat{a}^+$ and $\hat{a}$ is represented directly by the algebra of the complex variable $z$ and its derivative $\partial_z$, respectively the particle number $n$ by the product $\hat{a}^+a = \vartheta_{|z|^2}$.

To support the theoretical considerations presented above, we examined, as example, the quantum systems with a linear energy spectrum.

**Appendix A 1 – Deduction of Eq. (3.40)**

$$\int_0^{R_c \leq \infty} d(|z|^2) \, G_{p,q+1}^{q+1,0}\left(|z|^2 \left| \begin{array}{c} / \ ; \ a-1 \\ 0, \ b-1 \ ; \ / \end{array}\right.\right) {}_rF_s(c; d; C|z|^2) =$$

$$= \sum_{n=0}^{\infty} \frac{1}{\rho_{r,s}(d/c\,|\,n)} C^n \int_0^{R_c \leq \infty} d(|z|^2) \, G_{p,q+1}^{q+1,0}\left(|z|^2 \left| \begin{array}{c} / \ ; \ a-1 \\ 0, \ b-1 \ ; \ / \end{array}\right.\right) (|z|^2)^n =$$

$$= \frac{\prod_{j=1}^{q}\Gamma(b_j)}{\prod_{i=1}^{p}\Gamma(a_i)} \sum_{n=0}^{\infty} \frac{\rho_{p,q}(b/a\,|\,n)}{\rho_{r,s}(d/c\,|\,n)} C^n = \frac{\prod_{j=1}^{q}\Gamma(b_j)}{\prod_{i=1}^{p}\Gamma(a_i)} \sum_{n=0}^{\infty} (1)_n \frac{\prod_{j=1}^{q}(b_j)_n \prod_{i=1}^{r}(c_i)_n}{\prod_{i=1}^{p}(a_i)_n \prod_{j=1}^{s}(d_j)_n} \frac{C^n}{n!} = \quad (3.40)$$

$$= \frac{\prod_{j=1}^{q}\Gamma(b_j)}{\prod_{i=1}^{p}\Gamma(a_i)} {}_{q+r+1}F_{p+s}(1, b, c; a, d; C)$$

**Appendix A 2 – Deduction of Eq. (4.43)**

$${}_1F_1(1; e_0+1; |z|^2) \, {}_0F_0(\,;\,;(e^{-\beta\hbar\omega}-1)|z|^2) = {}_1F_1(1; e_0+1; |z|^2 \exp[(e^{-\beta\hbar\omega}-1)|z|^2]) =$$

$$= \sum_{m=0}^{\infty} \frac{(1)_m}{(e_0+1)_m} {}_2F_1(-m, 1-m-e_0-1; 1-m-1; (1-e^{-\beta\hbar\omega})) \frac{(|z|^2)^m}{m!} =$$

$$= \sum_{m=0}^{\infty} \frac{(1)_m}{(e_0+1)_m} {}_1F_0(-m-e_0; \, ; (1-e^{-\beta\hbar\omega})) \frac{(|z|^2)^m}{m!} = \quad (4.43)$$

$$= \sum_{m=0}^{\infty} \frac{(1)_m}{(e_0+1)_m} (e^{-\beta\hbar\omega})^{m+e_0} \frac{(|z|^2)^m}{m!} = e^{-\beta\hbar\omega e_0} \sum_{m=0}^{\infty} \frac{(1)_m}{(e_0+1)_m} \frac{(e^{-\beta\hbar\omega}|z|^2)^m}{m!} =$$

$$= e^{-\beta\hbar\omega e_0} \, {}_1F_1(1; e_0+1; e^{-\beta\hbar\omega}z|^2)$$